\documentclass[a4paper,11pt]{article}
\usepackage{jinstpub} % for details on the use of the package, please see the JINST-author-manual
\usepackage{lineno}
\usepackage{subfigure}
\usepackage{multirow}
\usepackage{threeparttable}
\title{\boldmath Optimization of WLS fiber readout for the HERD calorimeter}

\author[a,b,c,1]{X. Liu,\note{Corresponding author.}}
\author[a,b,2]{Z. Quan,\note{Corresponding author.}}
\author[a,b,3]{Y.W. Dong,\note{Corresponding author.}}
\author[a,b]{M. Xu}
\author[a,b]{J.J. Wang}
\author[a,b]{R.J. Wang}
\author[a,b]{Z.G. Wang}
\author[a,b]{X.Z. Cui}
\author[a,b]{T.W. Bao}
\author[a,b,c]{C.L. Liao}
\author[d]{J.F. Han} 
\author[a,b]{Y. Chen}

\affiliation[a]{Institute of High Energy Physics, Chinese Academy of Sciences,\\No. 19B, Yuquan Road, Shijingshan District, Beijing, China}
\affiliation[b]{Key laboratory of particle astrophysics, Chinese Academy of Sciences,\\No. 19B, Yuquan Road, Shijingshan District, Beijing, China}
\affiliation[c]{University of Chinese Academy of Sciences,\\No.1 Yanqihu East Rd, Huairou District, Beijing, China}
\affiliation[d]{Sichuan University,\\No.24 South Section 1, Yihuan Road, Chengdu , China}

\emailAdd{xliu@ihep.ac.cn; quanzheng@ihep.ac.cn; dongyw@ihep.ac.cn}

\abstract{A novel 3-D calorimeter, composed of about 7500 LYSO cubes, is the key and crucial detector of the High Energy cosmic-Radiation Detection (HERD) facility to be installed onboard the China Space Station. Energy deposition from cosmic ray in each LYSO cube is translated by multiple wavelength shifting (WLS) fibers for multi-range data acquisition and real-time triggering.

In this study, various methods of surface finish and encapsulation of the LYSO cube were investigated to optimize the amplitude from the WLS fiber end with the aim of improving the signal-to-noise ratio of Intensified scientific CMOS (IsCMOS) collection. The LYSO cube with five rough surfaces and a specular reflector achieves the maximum amplitude at the low-range fiber end, which is increased by roughly 44\% compared to the polished cube with PTFE wrapping.

The non-uniformity of amplitude at different positions on the LYSO cube surface was measured by X-ray and the positional correlation factor was derived for the entire cube. A simulation based on HERD CALO was conducted, which revealed that both the LYSO cube with five rough surfaces and the cube with rough bottom face exhibit superior energy resolution for electrons compared to the other two configurations.

}

\keywords{Calorimeters; Scintillators, scintillation and light emission processes (solid, gas and
liquid scintillators); Detector modelling and simulations I (interaction of radiation with matter, interaction of photons with matter, interaction of hadrons with matter, etc)}

\arxivnumber{1234.56789} % Only if you have one

\begin{document}
\maketitle
\flushbottom

\section{Introduction}

The High Energy cosmic-Radiation Detection (HERD) facility has been proposed as a space astronomy and particle physics experiment that will be installed on the China Space Station around 2027. The maximum size of the envelope in orbit is approximately 4.3×2.6×1.9 m³, with a weight of around 4400 kg.

Its primary scientific goals are to search for dark matter signals in the energy spectra and anisotropy of high-energy electrons and gamma-rays, and to measure precisely the energy spectra and composition of primary cosmic rays up to the knee energy, as well as to monitor the high-energy
gamma-ray sky.\cite{Zhang:2017} \cite{Dong}\cite{xu2016}\cite{Gargano},

The HERD payload consists of a calorimeter (CALO)\cite{Pacini:2021kD}, a fiber tracker (FIT)\cite{Perrina}, a plastic scintillator detector (PSD)\cite{Kyratzis}, a silicon charge detector (SCD) and a transition radiation detector (TRD)\cite{Liu2020}. The CALO is the core detector of the HERD, and it is a homogeneous, 3D segmented calorimeter made up of about 7500 LYSO cubes with 3 cm side lengths, corresponding to roughly 55 radiation lengths(X0) and 3 nuclear interaction lengths for centrally incident particles in any directions, respectively. 

LYSO is the good candidate for space high energy calorimeters due to its high light yield (30 photons per keV), fast decay time (approximately 37 ns), long attenuation length (1.2 cm for 511 keV), and low average temperature coefficient (-0.28\%/$^\circ C$  from 25$^\circ C$ to 50$^\circ C$)\cite{LYSO}.

Each LYSO cube’s fluorescence is read out in two independent approaches: the first approach uses wavelength-shifting (WLS) fiber to deliver the light to Intensified scientific CMOS (IsCMOS) cameras, the second approach uses photodiode (PD) sensors\cite{Adriani_2022}. The reliability of the CALO in-orbit data acquisition can be significantly improved by cross-calibration between the double read-out systems.
 
Two WLS fibers are attached to a face of each cube, and both ends of each WLS fiber are utilized for fluorescence read-out. Two fiber ends are connected to a high-range IsCMOS and a low-range IsCMOS respectively, while the other two ends are linked to a trigger system.
Each IsCMOS consists of an Image Intensifier(II) followed by a scientific CMOS camera and has a 5000 times dynamic range. The light emitted from the WLS fiber is focused on a photocathode through a front taper in II, where it is converted into electrons. These electrons are multiplified on a microchannel plate (MCP) and are subsequently reconverted to photons via phosphor. Finally, these photons are focused on the CMOS camera by a rear taper\cite{Pacini:2021kD}.

The two IsCMOS have different electron multiplication capabilities in their MCPs, with the IsCMOS featuring relatively high magnification serving as the low range and the other camera serving as the high range. By properly selecting specific gain of MCPs, the two IsCMOS combined can achieve more than $10^7$ times dynamic range. The phosphor's decay time is approximately 200$u$s, while the CMOS has a high frame rate (>800 frames per second) and low readout noise (<1.5 electrons)\cite{Pacini:2021kD}. 

The ﬁbers dedicated to the trigger systems are grouped to couple with several Photo Multiplier Tubes(PMT) which provide fast trigger information regarding the energy deposition in speciﬁc regions. The read out of both ﬁber ends can backup each other within the trigger system.

\begin{figure}[ht]
\centering
\includegraphics[width=3in]{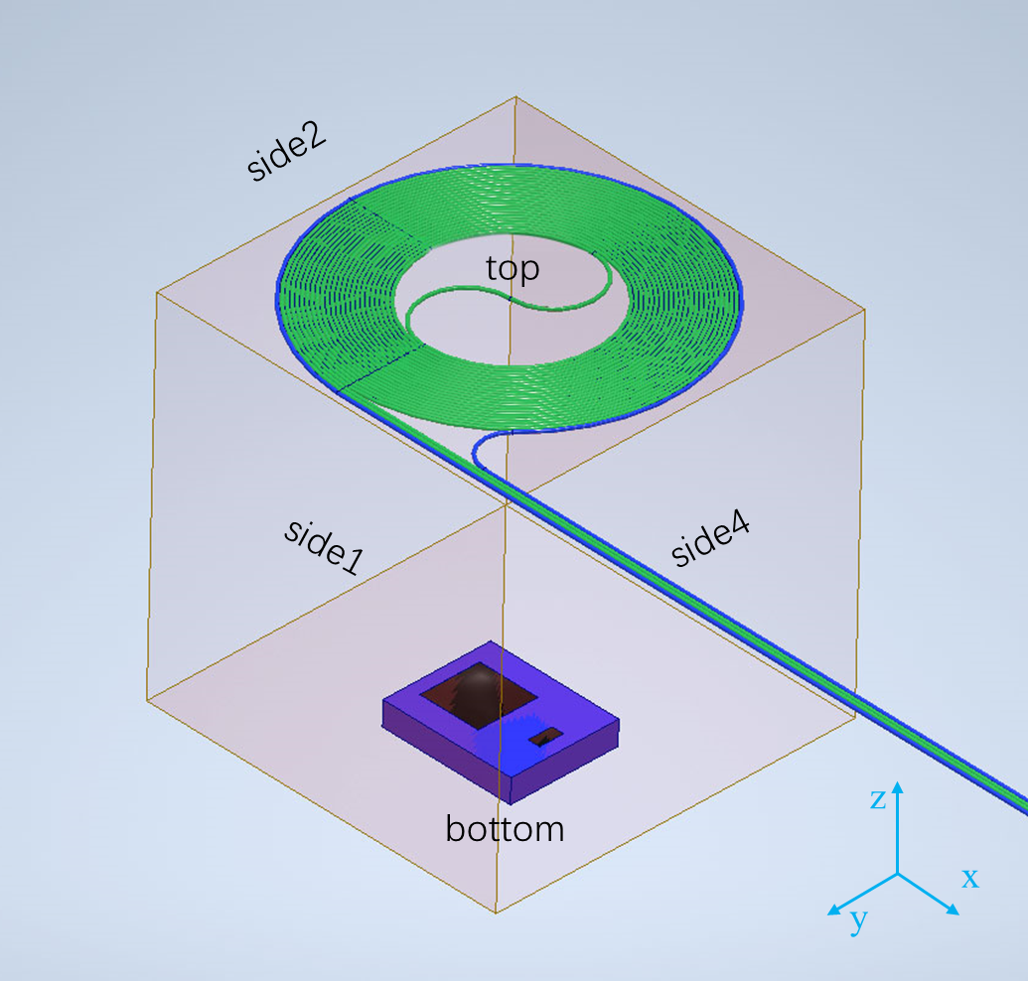}
\caption{Illustration of the LYSO cube and its read-out. Side 3 is opposite to side 1. The green piece denotes the inner fiber, while the blue piece represents the outer one. A pair of PDs are affixed to the bottom surface of the cube.}\label{fig:cube}
\end{figure}

The lowest detection limit of LYSO requiring a corresponding large enough amplitude at the fiber end coupling to low-range IsCMOS. To optimize the amplitude distribution of the four WLS fiber ends, two 0.3 mm diameter WLS fibers are wound in different loops. The inner fiber (Kuraray Y-11(200)MS) is wound 11 loops, while the outer fiber (Kuraray Y-11(100)MS) makes one loop around the inner one, as illustrated in Figure \ref{fig:cube}. The length of the fiber outside LYSO is 2.4 meters, ensuring connectivity to the fiber at the farthest position relative to IsCMOS and consistent light output across all fibers within their respective ranges. The inner fiber end exhibits a larger contact area with LYSO, resulting in a higher light output compared to the outer one. The ratio of light output between the inner and outer fiber ends is approximately 15:1. 

In order to achieve a good signal-to-noise (S/N) ratio, one of the inner fiber ends is selected as the low-range fiber. The high-range IsCMOS is designed to operate with lower light levels than the low-range IsCMOS in order to prevent saturation, so an outer fiber is connected to it as the high-range fiber. The remaining two fiber ends are routed to the trigger system.

The winding part of fibers are coated with silicone elastomer (DOW CORNING Sylgard 184) and molded into a mat of 0.8 mm thickness to enhance coupling with the LYSO top face. A monolithic package, comprising a Large PD (LPD) with an active area of 25 mm$^2$ and a small SPD with an active area of 1.6 mm$^2$, is affixed to the bottom face of the LYSO cube\cite{Adriani_2022}, as illustrated in Figure \ref{fig:cube}.The LPD-to-SPD signal ratio is approximately 1500:1, while both PD combinations exhibit dynamic ranges exceeding $10^7$times.
A reflector serves as a light shielding layer for each cube, enhancing both the light collection efficiency and the prevention of cross-talk between cubes.

The read-out amplitude of the WLS fiber significantly affects the S/N ratio collected by the IsCMOS, and the non-uniformity of the amplitude at different positions within the cube may impact the energy resolution of HERD CALO, which plays a key role in achieving the science goals, such as electron and gamma-ray observations.

This paper will present the improvement of read-out amplitude of the WLS fiber by optimizing the surface roughness and coating of the LYSO, the non-uniformity measurement results under different surface roughness and the simulation of its impact on energy resolution.

\section{Experimental setup}
\label{sec:setup}
\begin{figure}[ht]
\centering
\includegraphics[width=5in]{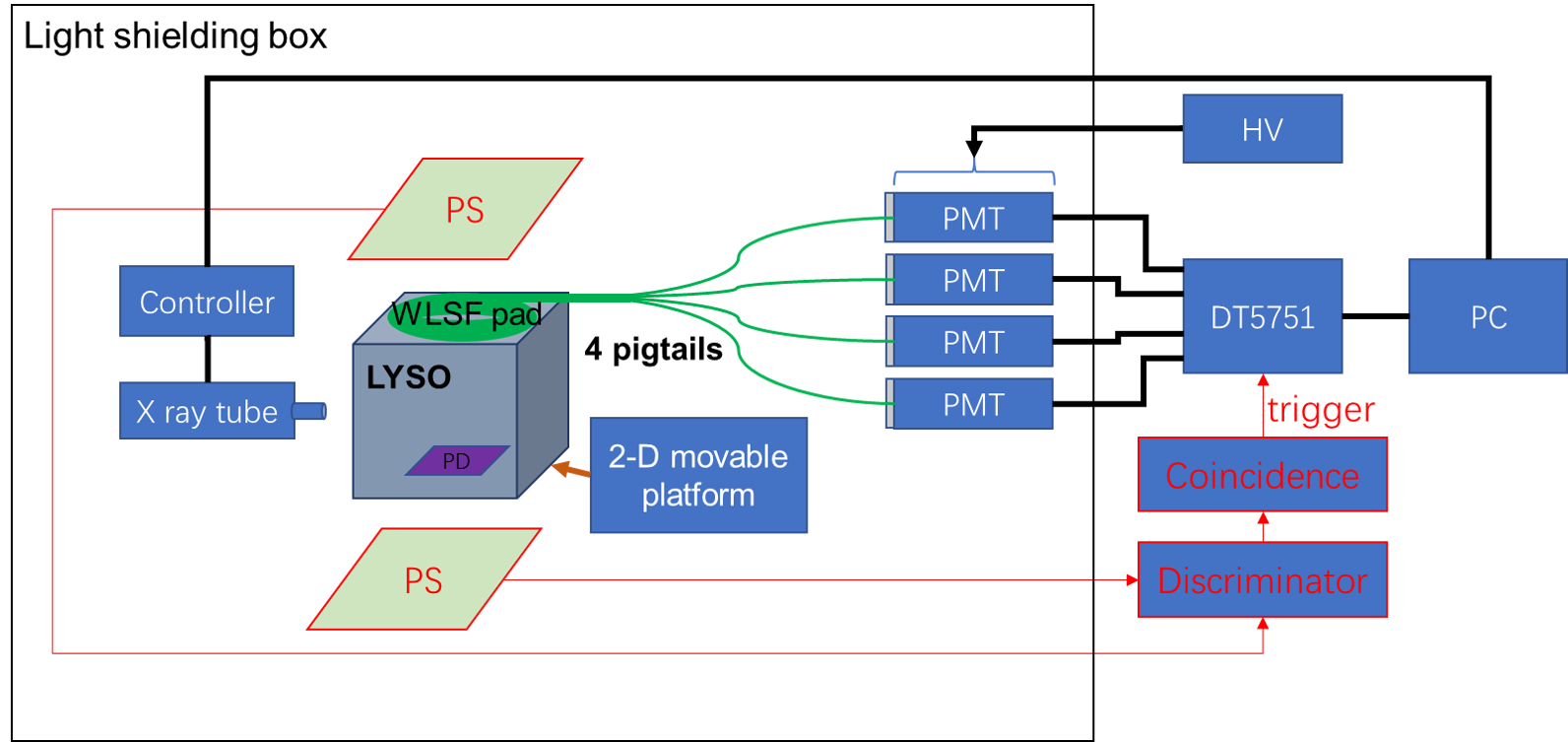}
\caption{The setup of measurement system.}\label{fig:setup}
\end{figure}

The measurement system consists of two plastic scintillation (PS) detectors, a discriminator (CAEN N840), a coincidence unit (CAEN N455), four photomultiplier tubes (Hamamatsu CR110), a high-voltage power supply (CAEN DT1470), a waveform sampling digitizer (CAEN DT5751), a computer, an X-ray tube with a silver target (Amptek mini x2)\cite{x2}  and a 2D movable platform, as shown in Figure \ref{fig:setup}. The detection area of each PS detector is 40 $\times$ 40 $mm^2$.

This system offers two acquisition modes through the loading of specific configurations. To measure minimum ionized particle (MIP) signals, the digitizer is configured in external triggering mode and receives triggers from PS detector coincidence events via logic units. When the X-ray tube is activated for scanning the LYSO surface, the digitizer is set to software-triggered mode. The corresponding waveforms are recorded at a sampling frequency of approximately 300 Hz with a sampling width of 5 $\mu$s.

To attenuate low-energy X-rays, a tungsten foil with a thickness of 0.025 mm and an aluminum foil with a thickness of 0.254 mm are positioned in front of the X-ray tube window\cite{x2}. A copper collimator with a diameter of 2 mm is installed adjacent to the filter to restrict the output cone angle. The high voltage and current settings for the tube are adjusted to 50 kV and 68 mA, respectively, in order to enhance high-energy photons while avoiding PMT saturation.
 
\begin{figure}[ht]
\centering
\includegraphics[width=4in]{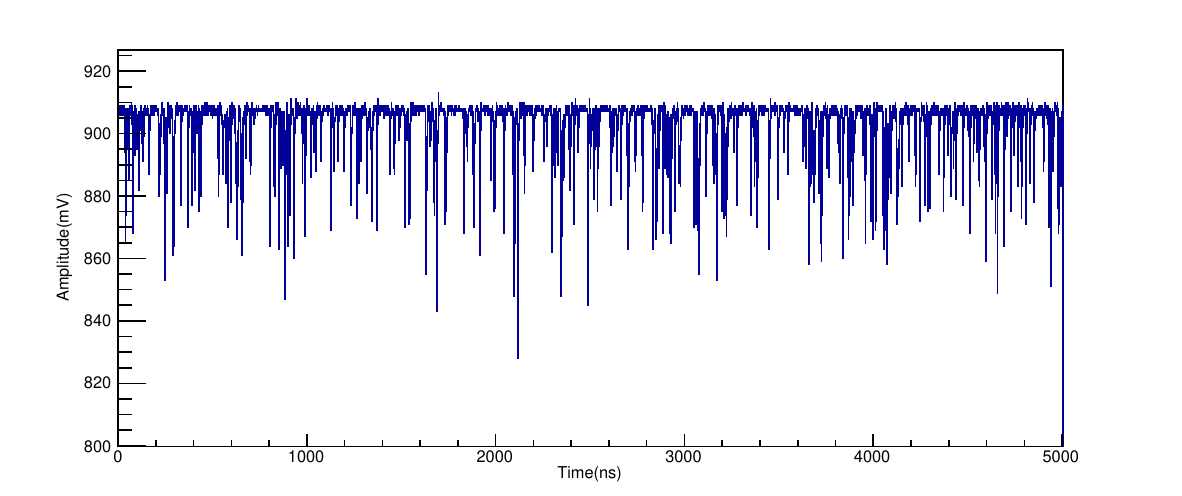}
\caption{Waveform in 5$\mu$s sampling time.}\label{fig:slice}
\end{figure}

\begin{figure}[ht]
\centering
\includegraphics[width=4in]{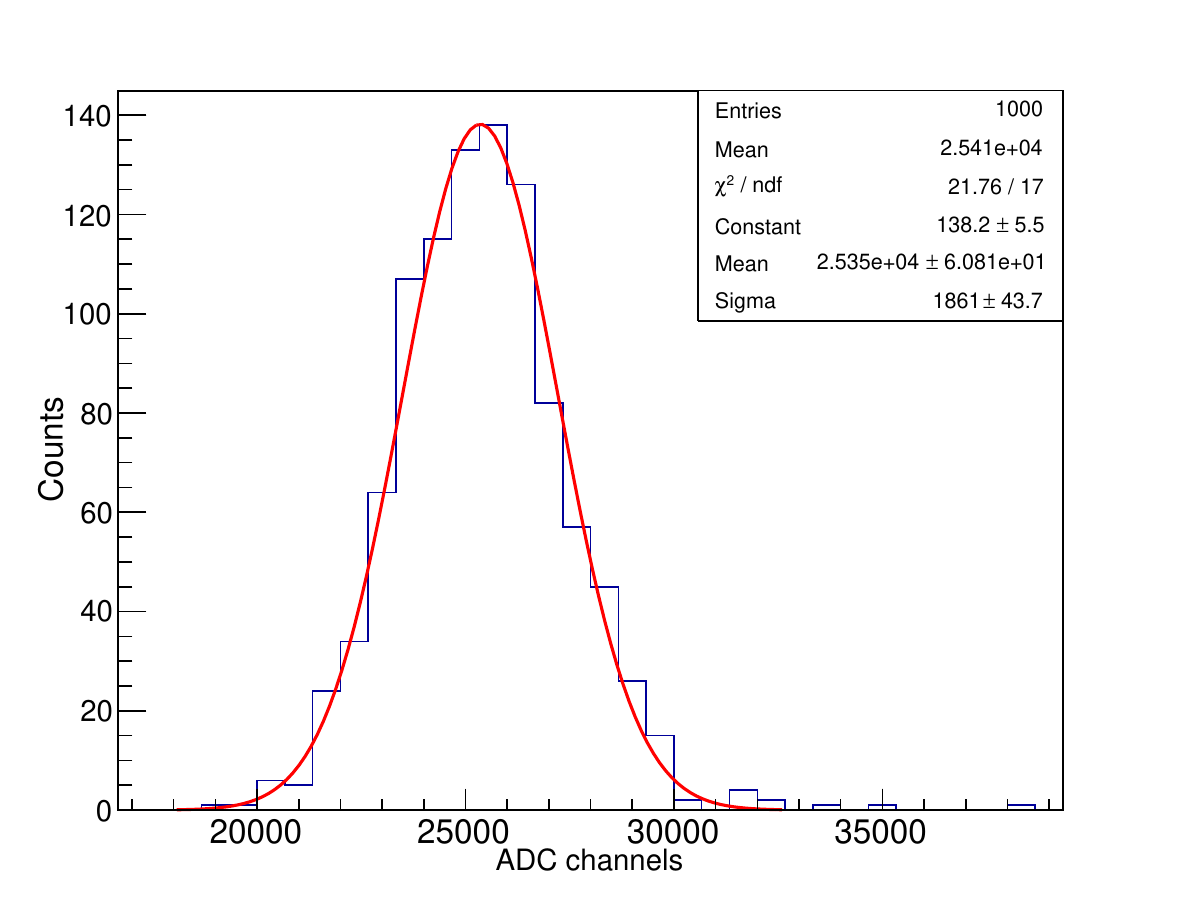}
\caption{Integral value distribution of 1000 waveforms.}\label{fig:slices}
\end{figure}

The low-range fiber produces approximately 10 photo-electrons(PE) (detected by a PMT with bi-alkali photocathode) per 1 MeV deposited in the LYSO cube. Therefore, when irradiated by a DC X-ray tube (with an upper energy limit of 50 keV), the PMT signal waveform comprises mainly of discrete single PE pulses, as illustrated in Figure \ref{fig:slice}, where the waveform consists of 181 PE pulses. Figure \ref{fig:slices} shows a corresponding distribution of integral values for 1000 waveform samples, with the mean value obtained from fitting a Gaussian function representing the signal amplitude at the end of the fiber.

\section{Measurement}
\subsection{LYSO surface finishing and coating}

This measurement employs the MIP signal, whose energy is sufficiently high to compensate for discrepancies in energy loss caused by different coating materials. Details of the measurement system can be found in section~\ref{sec:setup}. 

Two reflective materials were independently applied to cover LYSO cubes: one was an Enhanced Specular Reflector (ESR) from 3M, and the other was BC-642 Polytetrafluoroethylene (PTFE) Reflector Tape from Saint-Gobain.Both materials are representative of specular and diffuse reflective materials, respectively, and can be easily applied to the surface of batch scintillators. 
Titanium oxide, another common highly reflective material, was not included in this measurement due to its signal amplitude being only 50\% of the ESR film coating found in previous HERD prototype\cite{Dong_2017}. Many micro-structure machining techniques have been utilized to modify the surfaces of scintillator, demonstrating their potential in enhancing light output\cite{Slates}\cite{KNAPITSCH2011}\cite{Liu_2022}. However, practical application remains challenging due to processing complexity and inefficiency at this stage.

\begin{figure}[htbp]
\centering
\includegraphics[width=4in]{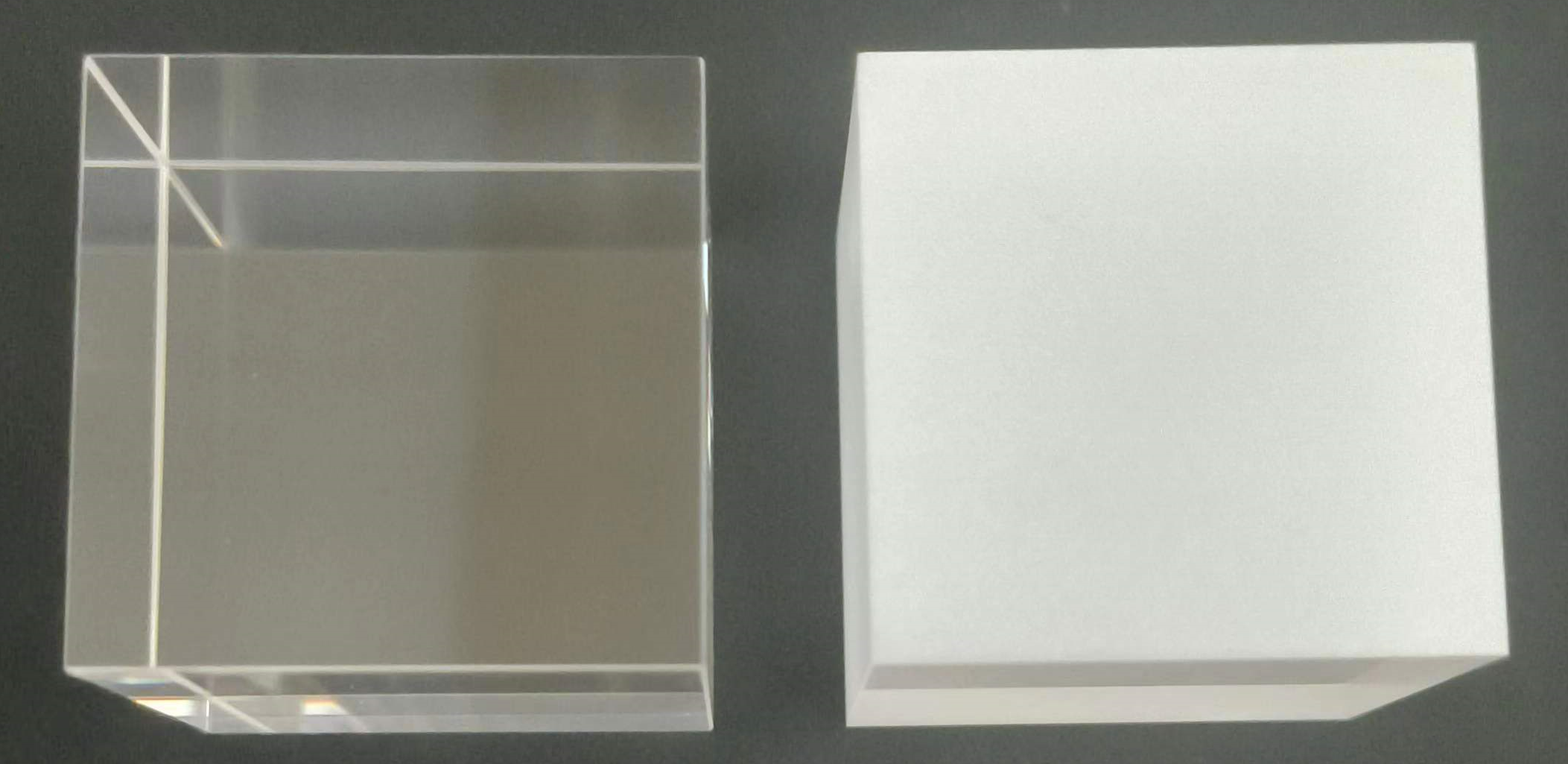}
\caption{Photo of two LYSO crystals. Left: LYSO with six polished faces; Right: LYSO with five rough facesand one polished face.}\label{fig:face}
\end{figure}

Two crystals with the same configuration were tested, and their surfaces (excluding the top surface) were polished or manually roughened using abrasive paper (STAR CHINA P400). All configurations are detailed in Table \ref{tab:config}. The PDs were not affixed to the bottom surface of LYSO to avoid being affected by non-uniform coupling. Figure \ref{fig:face} illustrates two LYSO crystals, one with six polished faces and the other with five rough faces and one polished face.

\begin{table}[]
\centering
\caption{Configurations of LYSO surface finishing and coating }
\begin{center}
\label{tab:config}
\begin{tabular}{|l|l|l|l|l|l|l|l|l|}
\hline
Config. number  & 1        & 2        & 3        & 4        & 5        & 6        & 7     & 8     \\ \hline
Reflector       & PTFE     & ESR      & PTFE    & ESR       & PTFE     & ESR      & PTFE  & ESR   \\ \hline
Bottom face     & polished & polished & rough    & rough    & polished & polished & rough & rough \\ \hline
Four side faces & polished & polished & polished & polished & rough    & rough    & rough & rough \\ \hline
\end{tabular}
\begin{tablenotes}
\item[1] Note: The top face is polished across all configurations.
\end{tablenotes}
\end{center}
\end{table}

\begin{figure}[htbp]
\centering
\includegraphics[width=4in]{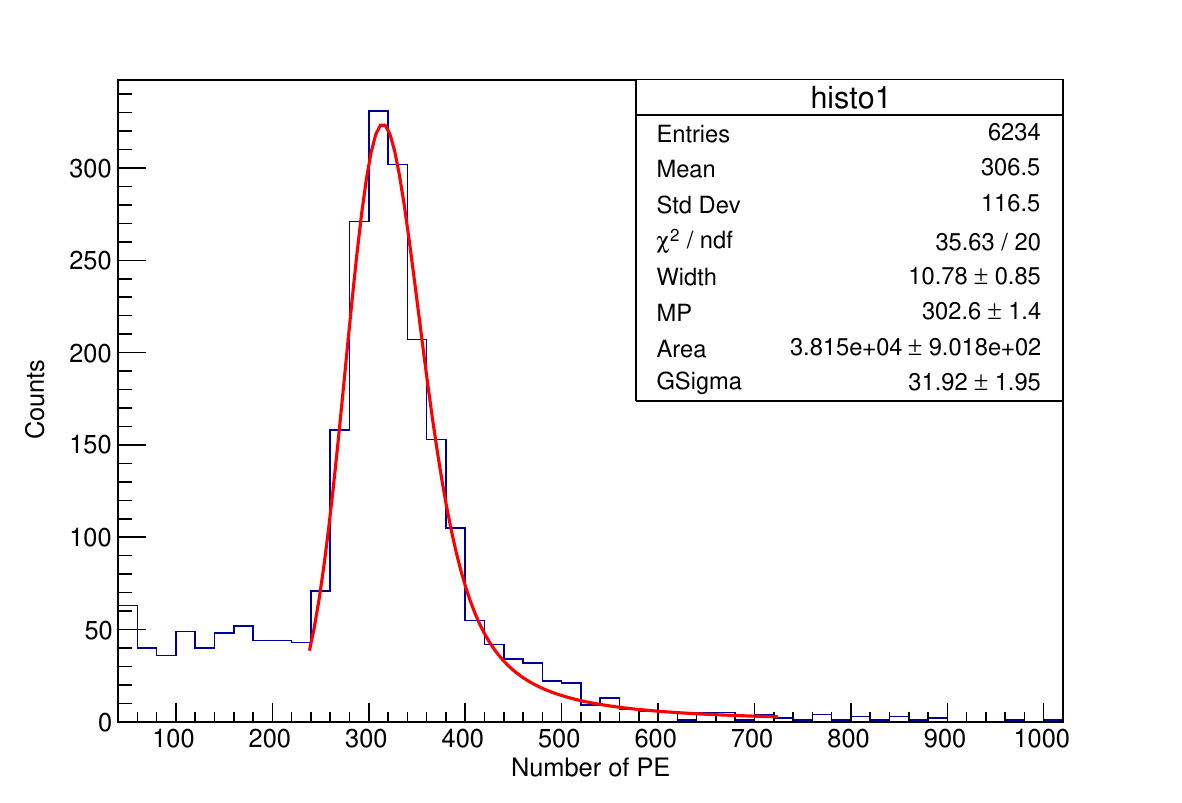}
\caption{A typical MIP signal spectrum.}\label{fig:mipspec}
\end{figure}

\begin{figure}[htbp]
\centering
\includegraphics[width=4in]{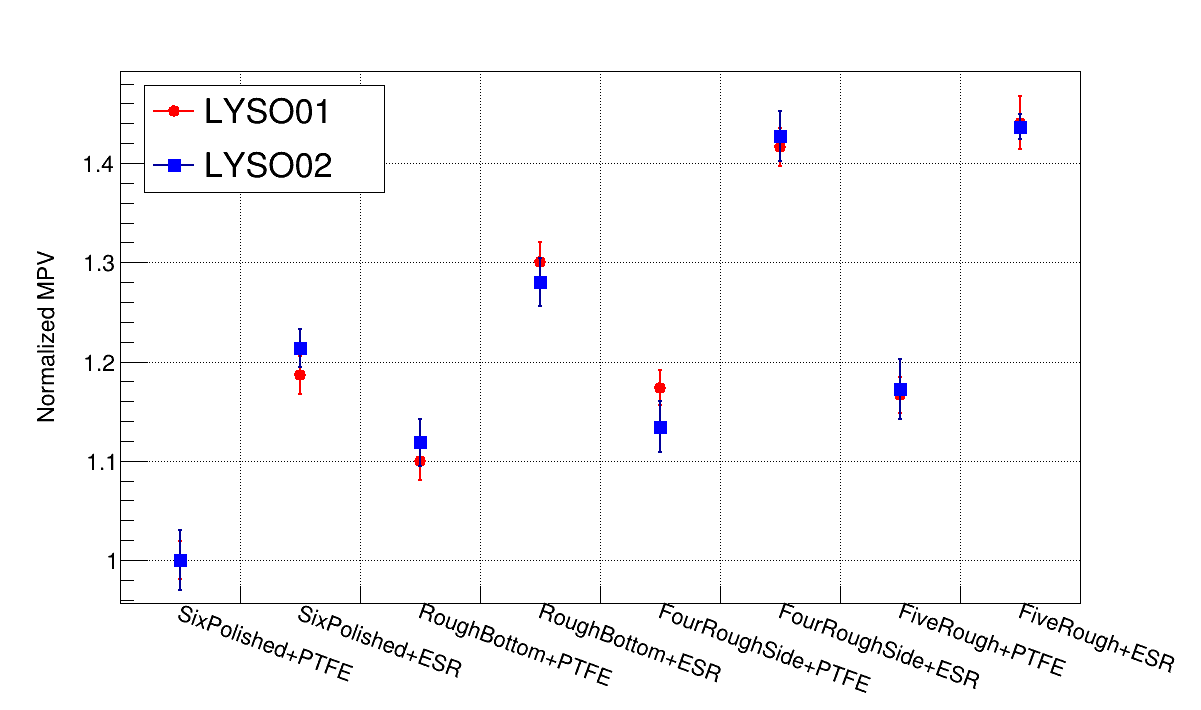}
\caption{The average normalized MPV values for various configurations. All MPV(listed in Table\ref{tab}) have been normalized to the minimum value among all configurations.The labels on the horizontal axis represent the surface conditions and coating materials of LYSO.}\label{fig:mpv}
\end{figure}

The amplitude distribution of the MIP signal from four WLS fiber ends with different configurations was obtained. A typical MIP signal spectrum fitted by Landau convoluted Gaussian function is showed in Figure \ref{fig:mipspec}, where the MP value(MPV) represents the most probable value obtained from fitting, and the horizontal axis scale has been converted to the number of PE. Table \ref{tab} presents the MPV of four WLS fiber ends across all configurations, while Figure \ref{fig:mpv} displays the corresponding average normalized  values relative to the minimum value among all configurations.

    \begin{table}[htbp]
    \centering
    \caption{MPV(in PE) of four WLS fiber ends belonging two LYSO across all configurations. The details of configuration can be found in Table \ref{tab:config}. Fiber 1 to Fiber 4 represent the four WLS fiber output ends. All data are converted to the number of PE.}
    \label{tab}
    \begin{tabular}{|l|l|l|l|l|l|l|l|l|}
    \hline
    \multicolumn{1}{|c|}{\begin{tabular}[c]{@{}c@{}}Config.\\ number\end{tabular}} & \begin{tabular}[c]{@{}l@{}}LYSO 01\\ Fiber 1\end{tabular} & \begin{tabular}[c]{@{}l@{}}LYSO 01\\ Fiber 2\end{tabular} & \begin{tabular}[c]{@{}l@{}}LYSO 01\\ Fiber 3\end{tabular} & \begin{tabular}[c]{@{}l@{}}LYSO 01\\ Fiber 4\end{tabular} & \begin{tabular}[c]{@{}l@{}}LYSO 02\\ Fiber 1\end{tabular} & \begin{tabular}[c]{@{}l@{}}LYSO 02\\ Fiber 2\end{tabular} & \begin{tabular}[c]{@{}l@{}}LYSO 02\\ Fiber 3\end{tabular} & \begin{tabular}[c]{@{}l@{}}LYSO 02\\ Fiber 4\end{tabular} \\ \hline
    1    & 261±1.8   & 258±1.1   & 13±0.2    & 13±0.1    & 238±1.6    & 231±1.3   & 11±0.2  & 11±0.2  \\ \hline
    2   & 309±1.9   & 307±1.9   & 15±0.2    & 15±0.2    & 288±1.7    & 280±1.5   & 14±0.2  & 14±0.2  \\ \hline
    3    & 285±2.0   & 284±2.1   & 14±0.2    & 14±0.2    & 265±2.0    & 259±2.1   & 13±0.2  & 13±0.2  \\ \hline
    4   & 338±2.1   & 335±2.2   & 17±0.2    & 17±0.2    & 305±2.2    & 297±2.2   & 15±0.2  & 15±0.2  \\ \hline
    5    & 301±1.9   & 298±1.4   & 15±0.2    & 15±0.2    & 267±2.4    & 266±1.7   & 13±0.2  & 13±0.2  \\ \hline
    6   & 366±1.9   & 360±1.9   & 19±0.2    & 18±0.3    & 329±2.7    & 326±2.6   & 17±0.3  & 17±0.2  \\ \hline
    7   & 296±1.9   & 291±2.0   & 15±0.2    & 16±0.2    & 278±1.9    & 272±1.8   & 13±0.3  & 13±0.3  \\ \hline
    8   & 368±3.8   & 364±2.8   & 19±0.3    & 19±0.4    & 333±1.3    & 327±1.5   & 17±0.1  & 17±0.2  \\ \hline
    \end{tabular}
    \end{table}

The two cubes were found to exhibit essentially same trend, where a cube with five rough faces and ESR coating had an signal amplitude increase roughly 44\% compared to a polished LYSO cube with PTFE wrapping. When comparing cubes with the same surface roughness, the cube coated with ESR film had a signal amplitude roughly 20\% higher than the one wrapped with PTFE, as shown in Table \ref{tab}. Thicker layers of PTFE wrapping were found to result in better performance\cite{Adriani_2019}, but would also increase CALO dead space leading to more energy leakage. Additionally, with the same coating material, the signal amplitude increased from polished to rough surfaces, which agrees with previous reports\cite{Adriani_2019}\cite{Ishibashi_1986}.

The impact of surface roughness and coating on light output can be explained by considering the transmission loss within the cube and the reflection loss at its surface. The rough surface serves as a diffuser, facilitating a shorter path for light transmission to the WLS fiber, thereby reducing both transmission losses and reflection frequencies. Moreover, ESR film exhibits higher reflection efficiency than PTFE.

\subsection{Non-uniformity of amplitude at different positions}
\begin{figure}[ht]
\centering
\includegraphics[width=3in]{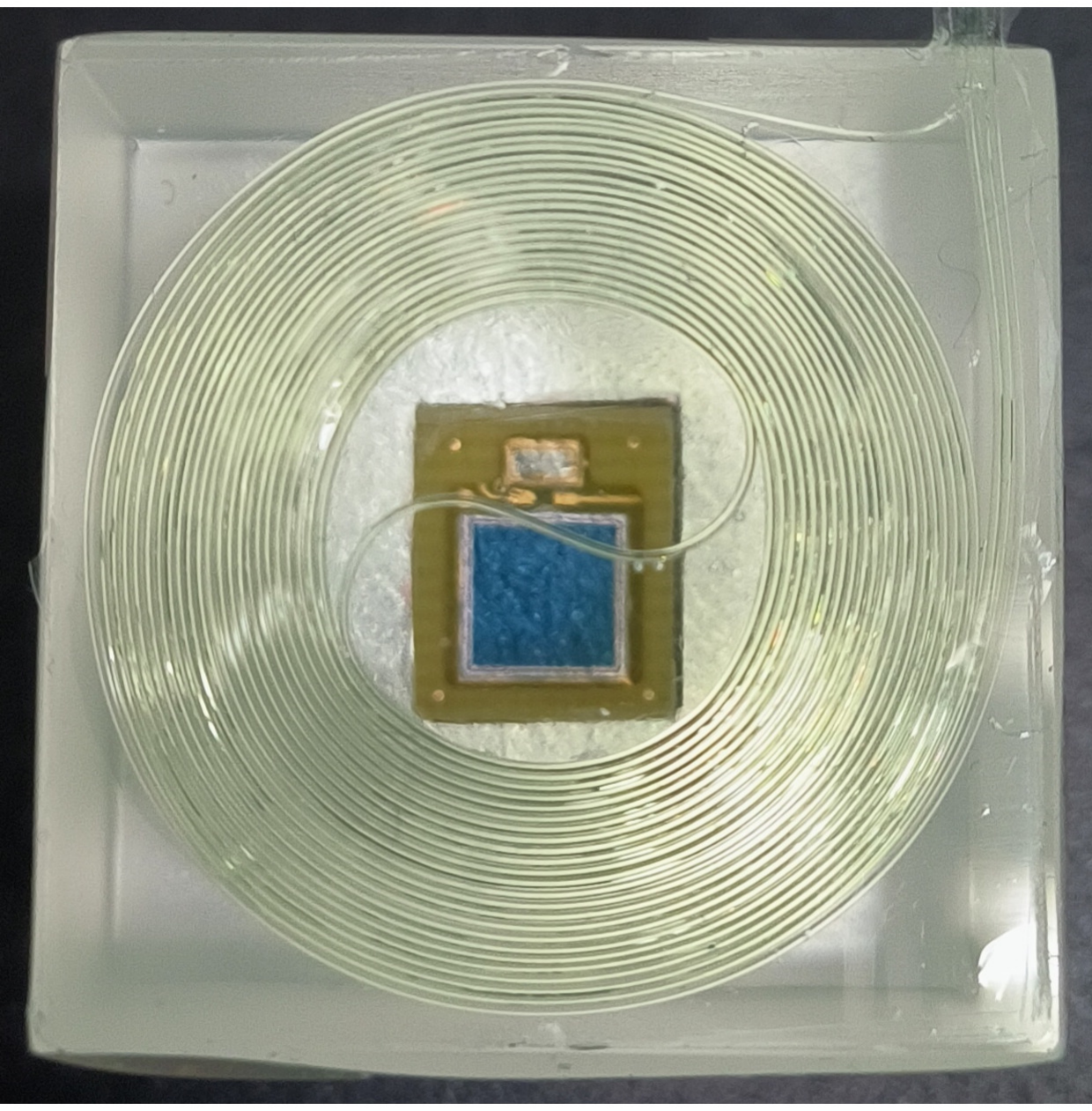}
\caption{Photo of LYSO cube with five rough faces. The ESR film covering WLS fiber mat is removed for clear visualization of the interior.}\label{fig:cubeP}
\end{figure}

To get the signal amplitude dependence on fired positions, the X-ray tube was utilized as an exciting source and scanned all six faces of a cube using a 2-D movable platform. The X-ray spot size taken on the cube surface was approximately 3 mm in diameter, with a scanning step set to 3 mm. 100 spots were measured on each face, covering almost the entire surface area. Four LYSO cubes with different surface roughness mentioned above were compared, utilizing only ESR film due to its higher light reflection efficiency and easier achievement of uniform thickness on all faces. Figure \ref{fig:cubeP} shows a photo of the cube with five rough faces and ESR coating.

\begin{figure}[ht]
\centering
\subfigure[A low-range fiber end]{
\includegraphics[width=5in]{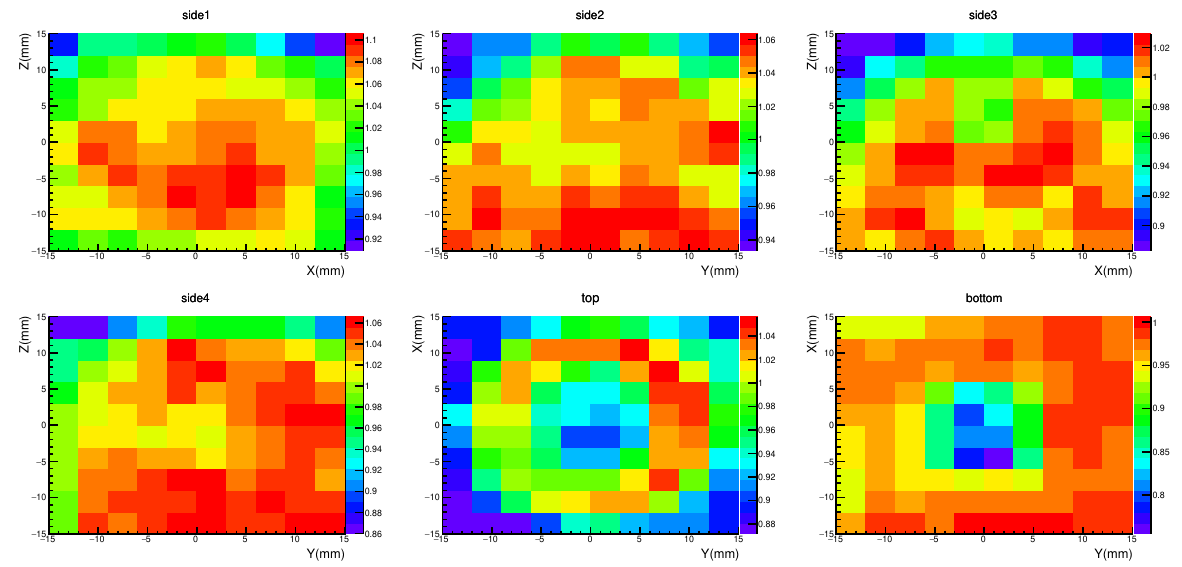}
\label{Figmapl}
}
\quad
\subfigure[A high-range fiber end]{
\includegraphics[width=5in]{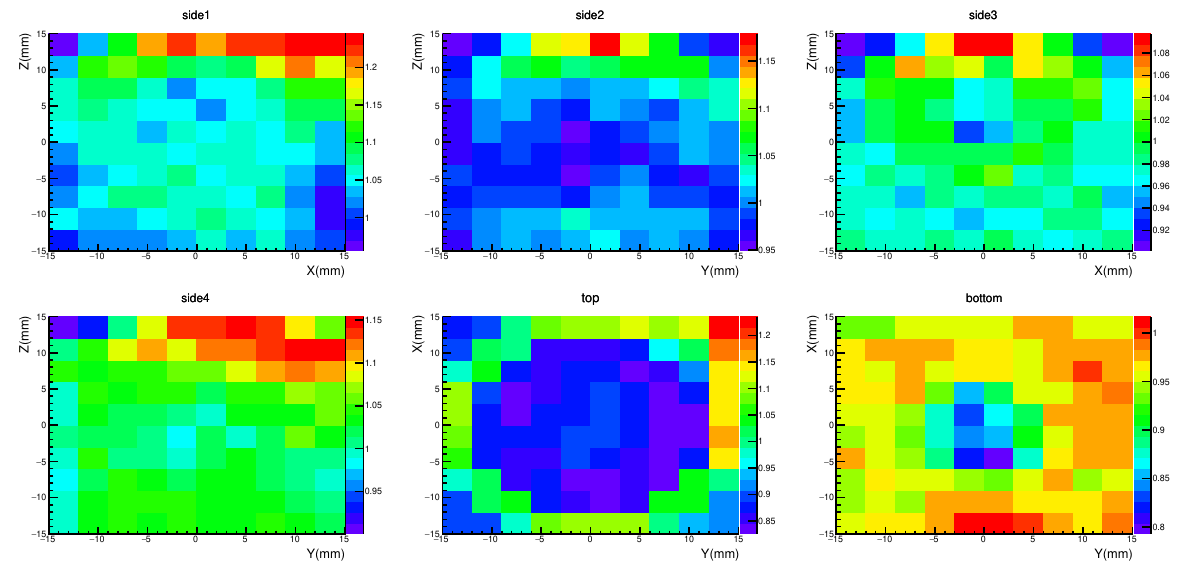}
\label{Figmaph}
}
\caption{A typical normalized signal amplitude at WLS fiber end dependence on X-ray excitation position.}\label{fig:map}
\end{figure}

Figure \ref{fig:cube} marks the definition of each face and coordinate axis, the typical amplitude variation map at low-range and high-range fiber end is showed in Figure \ref{fig:map}, where the attenuation caused by the WLS fiber mat and PD package was corrected. The data for each spot is normalized based on the average of all six faces' data. As seen in Figure \ref{fig:map}, the amplitude at one end of the fiber significantly increases when the exciting position is close to it, and both fiber ends' amplitudes are reduced at the excitation position close to PD. The performance of the two opposite ends of each fiber is essentially the same. The LYSO can absorb over 98\% of X-ray energy within less than 0.5 mm depth when exposed to maximum 50 keV X-rays, considering only a cover thickness of a 65um ESR film and a 100um KAPTON tape. So this test represents the most extreme case of non-uniformity of the cube. 

\begin{figure}[ht]
\centering
\subfigure[Low-range fiber]{
\includegraphics[width=2.5in]{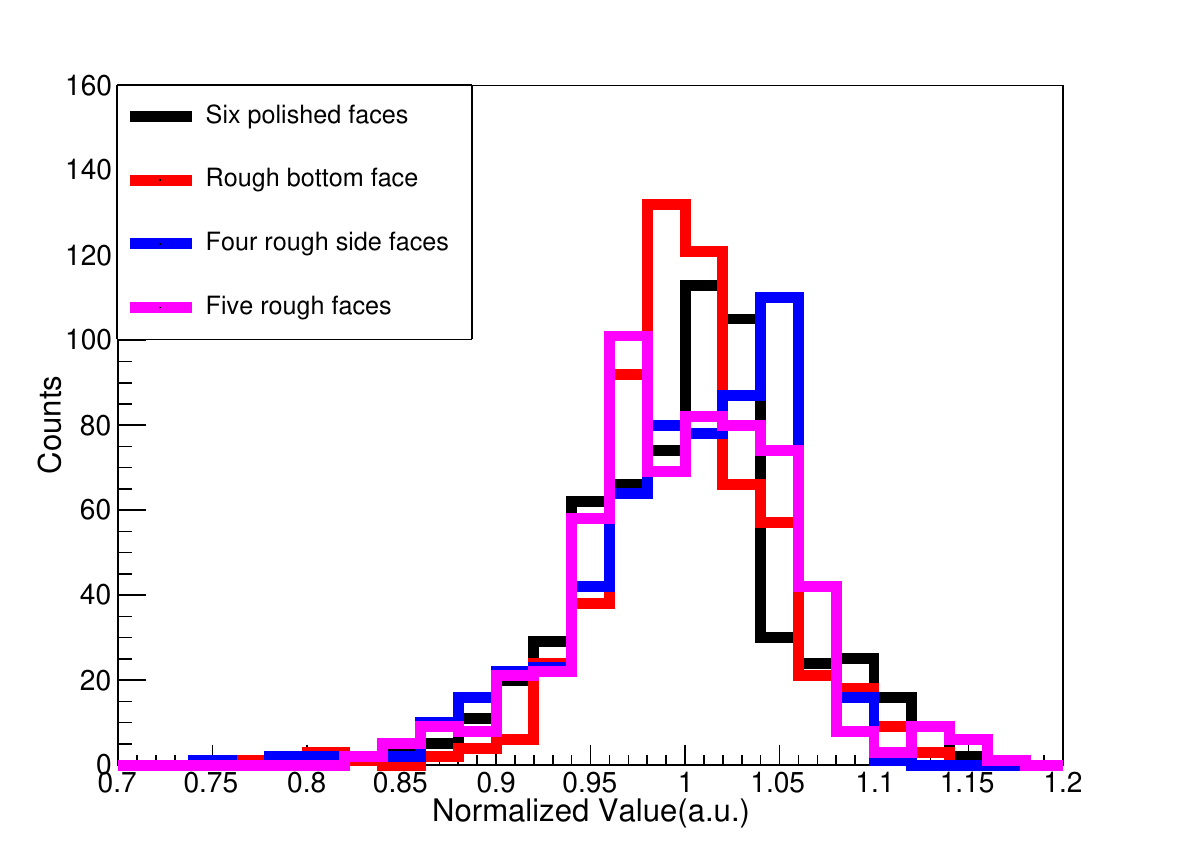}
\label{Fig2-1}
}
\quad
\subfigure[High-range fiber]{
\includegraphics[width=2.5in]{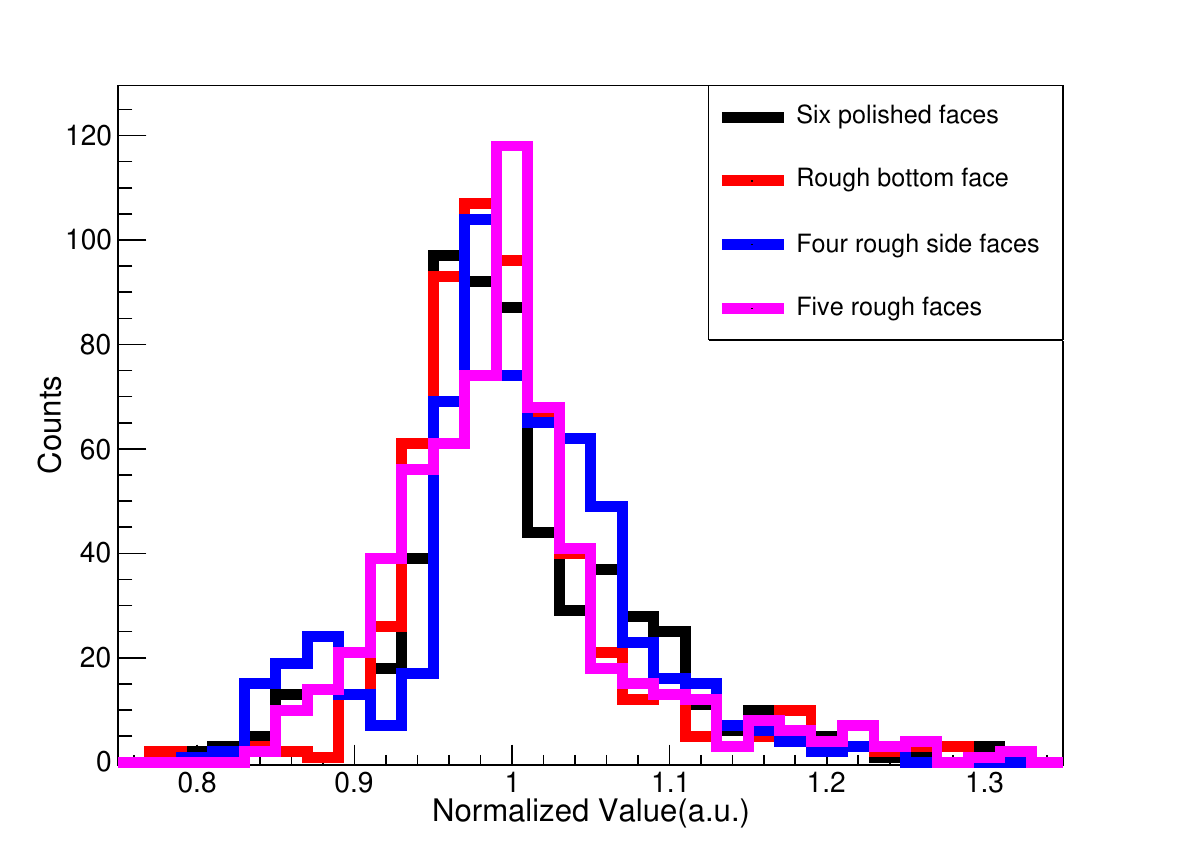}
\label{Fig2-2}
}
\caption{Distribution of fiber end output on X-ray scanning.}\label{fig:surface-distri}
\end{figure}

The distribution of normalized scanning results for different surface roughness is shown in Figure \ref{fig:surface-distri}. The corresponding standard deviation($\sigma$) and skewness($\gamma$) are listed in Table \ref{tab2}. The $\gamma$ characterizes the asymmetry of the distribution, which is derived from the corresponding histogram by the following equation:

\begin{equation}
\label{eq:x}
\gamma  =  \frac{\sum_{i=0}^{n} W_{i} \times  \left ( X_{i}  -\mu    \right ) ^{3}   }{N\times \sigma ^{3} } 
\end{equation}

Where  $n$ is the bin number,  $W$ is the bin content, $X$ is the bin center, $\mu$  is the mean value and $N$ is the number of entries. 

For low-range fibers, the smallest $\sigma$ is from the cube with rough bottom face. However, the cube with five rough faces has the smallest $\gamma$ value. For high-range fibers, the difference in $\sigma$ is smaller compared to low-range fibers, and the cube with six polished faces as well as the cube with four rough side faces have smaller $\gamma$ values than the other two configurations.

The map in Figure \ref{fig:map} demonstrates that the maximum light loss is due to PD absorption. However, roughening the bottom face can partially mitigate this effect, thereby improving non-uniformity. The roughing of all five faces enhances homogenization and promotes symmetrical amplitude distribution.

The contact area between high-range fibers and LYSO is small, resulting in low absorption efficiency and insensitivity to surface roughness changes. Additionally, there may exists a competitive relationship between high-range and low-range fibers in fluorescence absorption. Therefore, roughening the bottom face or five faces enhance the signal of low-range fibers but may have a negative impact on high-range fibers.

\begin{table}[htbp]
\centering
\caption{Statistical results($\sigma$ and $\gamma$) of the X-ray scanning for different surface roughness.}\label{tab2}
\begin{tabular}{|cc|c|c|c|c|}
\hline
\multicolumn{2}{|c|}{Surface condition}  & \begin{tabular}[c]{@{}c@{}}Six polished\\ faces\end{tabular} & \begin{tabular}[c]{@{}c@{}}Rough bottom\\ face\end{tabular} & \begin{tabular}[c]{@{}c@{}}Four rough \\ side faces\end{tabular} & \begin{tabular}[c]{@{}c@{}}Five rough\\ faces\end{tabular} \\ \hline
\multicolumn{1}{|c|}{\multirow{2}{*}{\begin{tabular}[c]{@{}c@{}}Low range\\ fiber\end{tabular}}}  & Std Dev  & 0.057 & 0.047 & 0.055  & 0.055    \\ \cline{2-6} 
\multicolumn{1}{|c|}{}  & Skewness & -0.312  & -0.696  & -0.946   & -0.095  \\ \hline
\multicolumn{1}{|c|}{\multirow{2}{*}{\begin{tabular}[c]{@{}c@{}}High range\\ fiber\end{tabular}}} & Std Dev  & 0.076     & 0.071   & 0.073    & 0.076       \\ \cline{2-6} 
\multicolumn{1}{|c|}{}    & Skewness & 0.741   & 1.333     & 0.105     & 1.191     \\ \hline
\end{tabular}
\end{table}

\section{Impact of signal amplitude non-uniformity on energy resolution}

\begin{figure}[ht]
\centering
\subfigure[Low-range fiber]{
\includegraphics[width=2.5in]{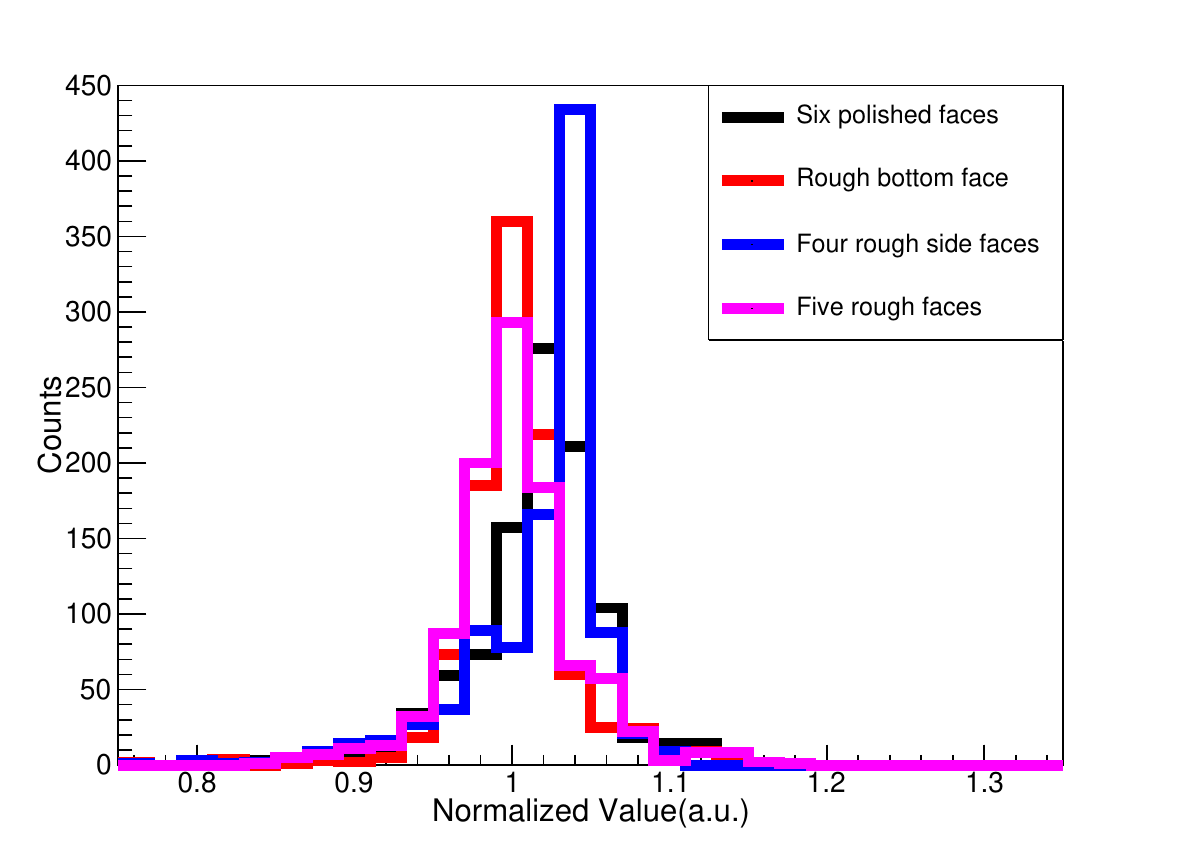}
\label{Fig2-3}
}
\quad
\subfigure[High-range fiber]{
\includegraphics[width=2.5in]{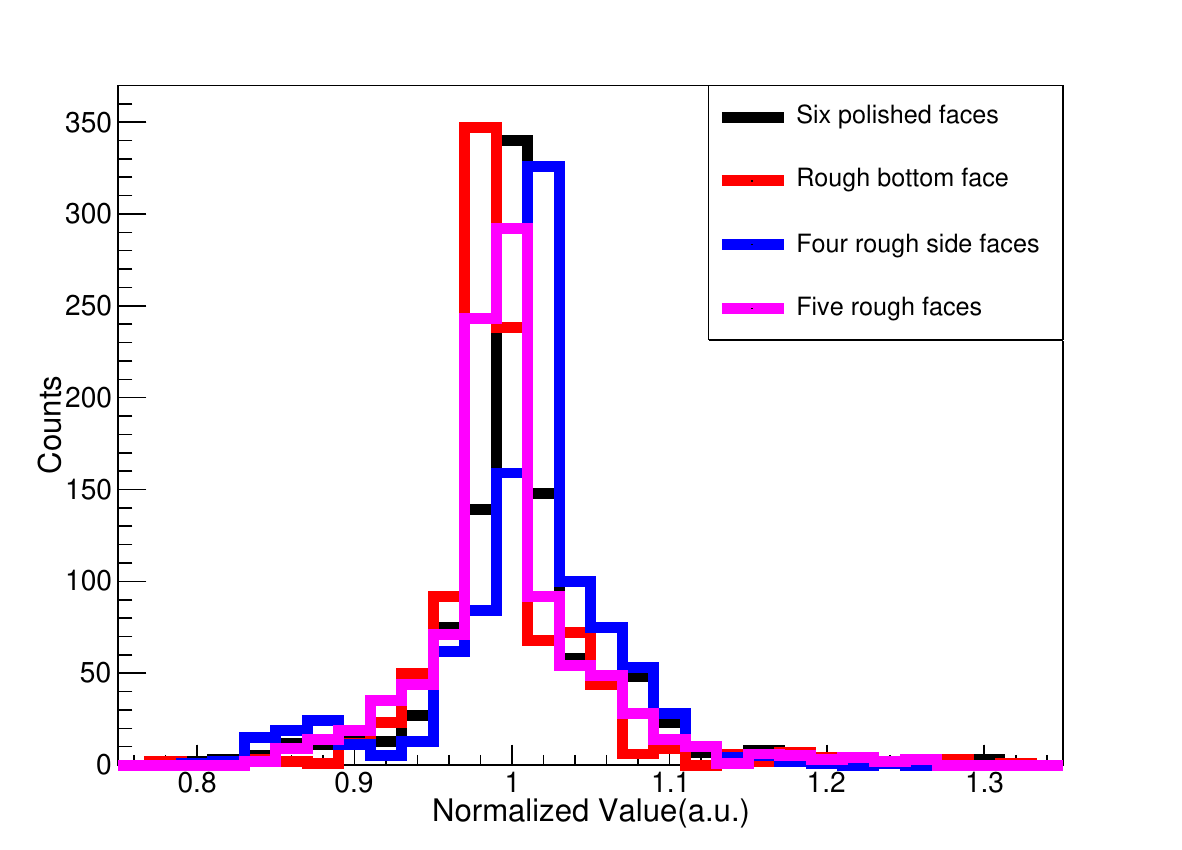}
\label{Fig2-4}
}

\caption{Derived distribution of fiber end output based on X-ray scanning data.}\label{fig:cube-distri}
\end{figure}

Based on surface scanning data, a positional correlation factor matrix that covers the entire LYSO cube was generated. In details, the cube was divided into 3 mm cubic units, and all normalized measurement data from surface scanning were used as the response weight of outermost layer of the cube, except for duplicate positions data, where the average value was adopted. The internal response of the cube was derived as the average of the data projected to the four lateral faces. The corresponding distributions are showed in Figure \ref{fig:cube-distri}.

Energy deposition in HERD CALO was simulated by extrapolating the positional correlation factor from a 2-D map of data. A crystal array geometry model consisting of 21 $\times$ 21 $\times$ 21 LYSO cubes was built in Geant4.10.7.2, and an isotropic electron particle gun with energies ranging from 10 to 1000 GeV was set up. For deposition energies less than 150 GeV in a single cube, the low-range fiber output was applied, while for higher energies, the high-range fiber was employed to simulate the practical high/low range read-out configurations. To emphasize the impact of signal amplitude non-uniformity and eliminate the influence of other factors, dead layer and crystal light yield are not incorporated in the simulation.

\begin{figure}[ht]
\centering
\subfigure[Six polished surfaces]{
\includegraphics[width=2.5in]{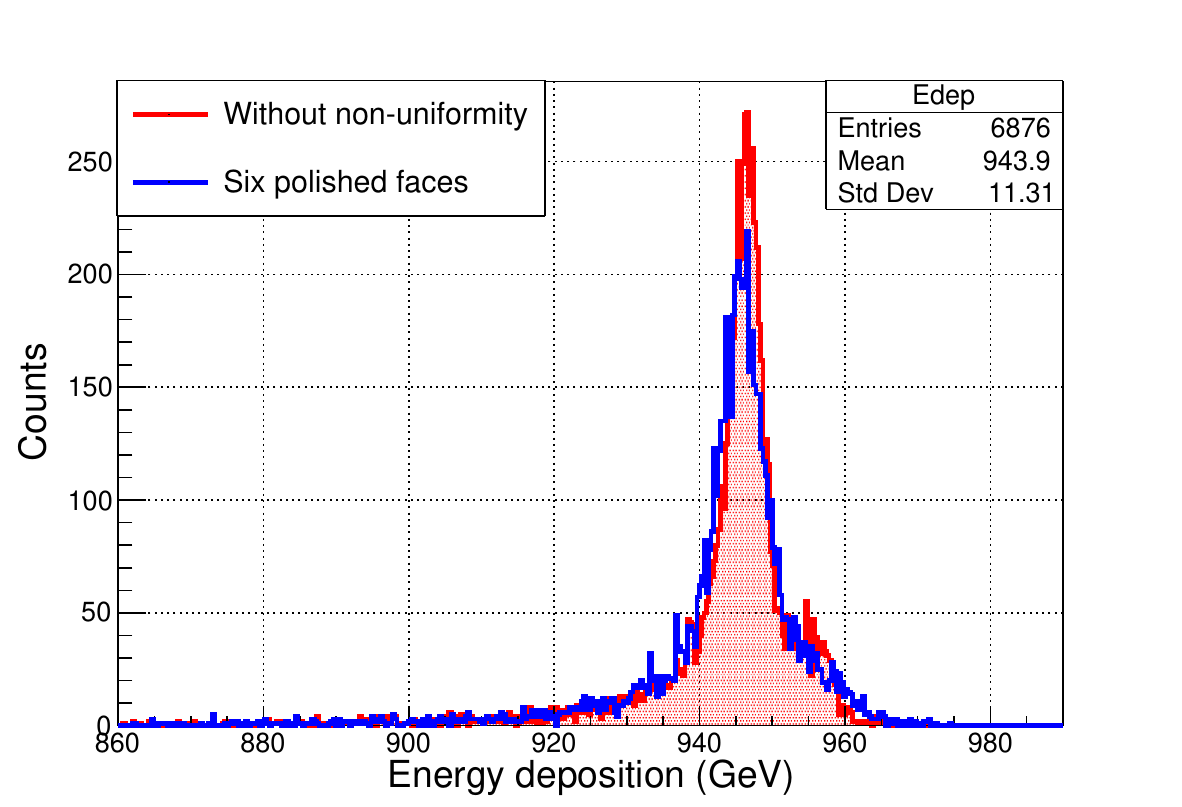}
\label{Fig3-1}
}
\quad
\subfigure[Rough bottom surface]{
\includegraphics[width=2.5in]{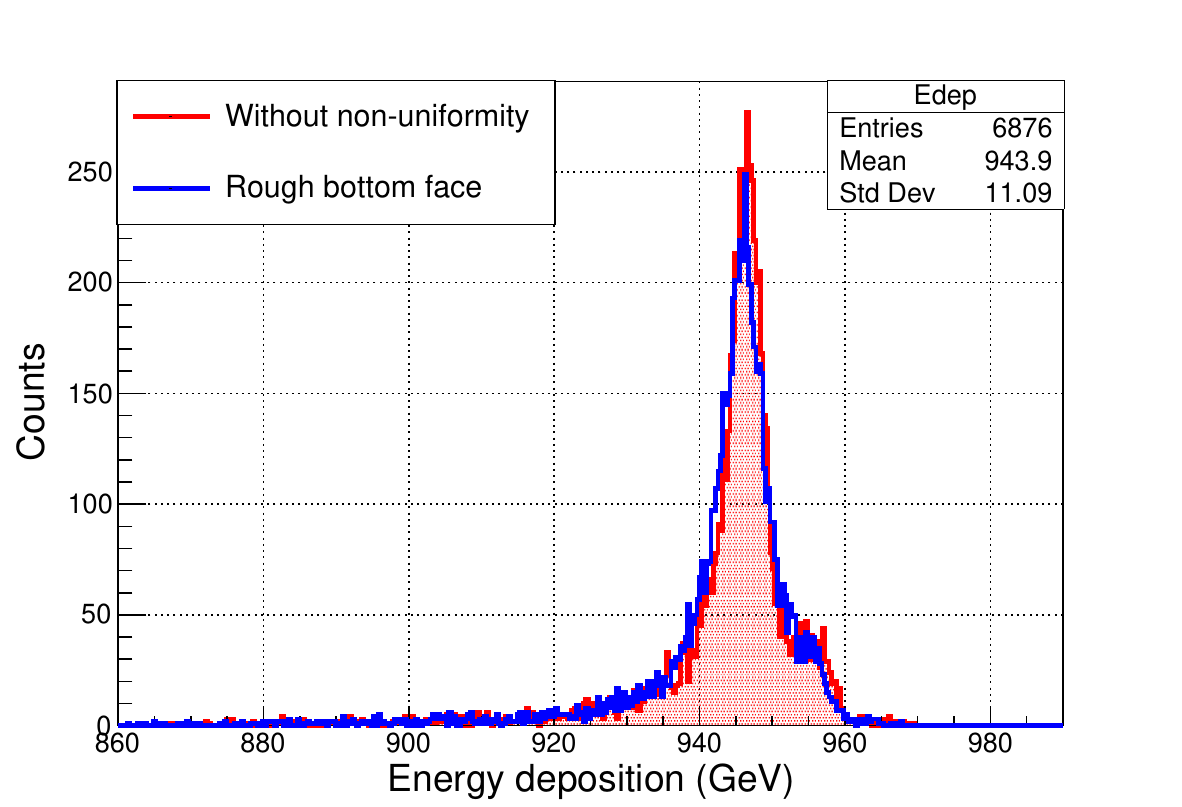}
\label{Fig3-2}
}
\quad
\subfigure[Four rough side surfaces]{
\includegraphics[width=2.5in]{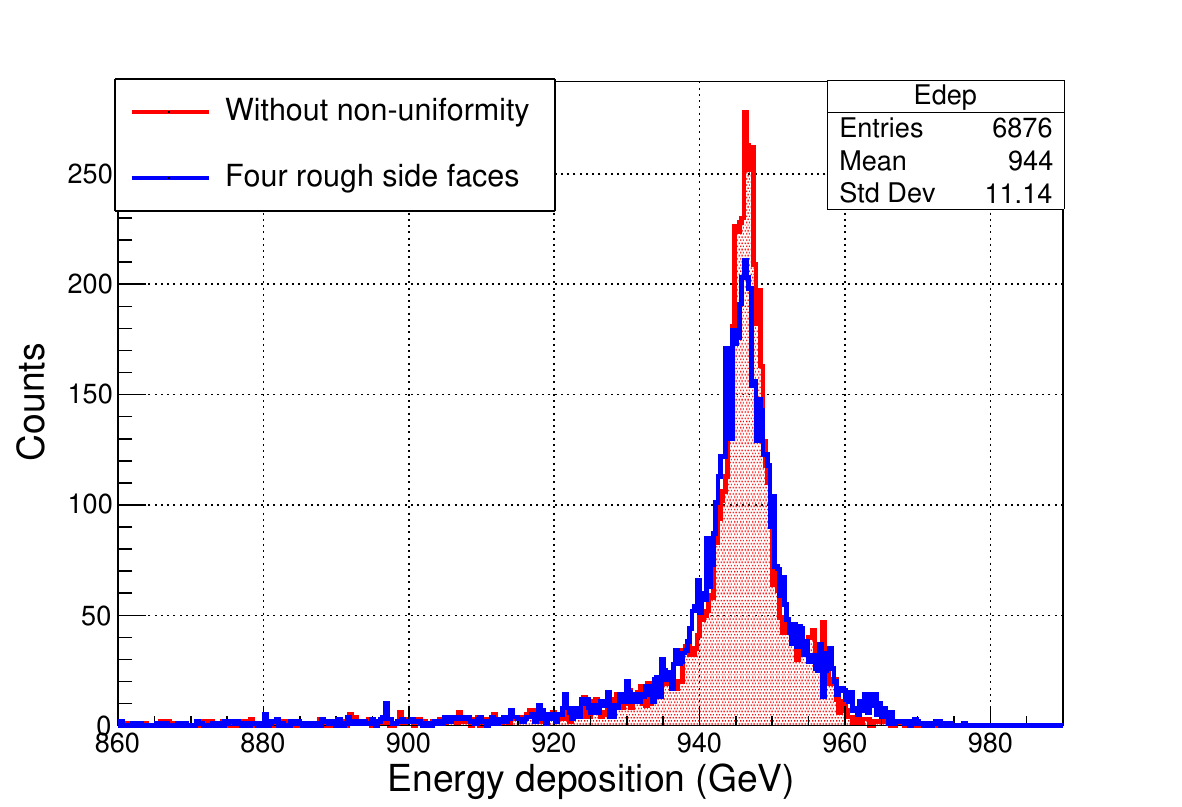}
\label{Fig3-3}
}
\quad
\subfigure[Five rough surfaces]{
\includegraphics[width=2.5in]{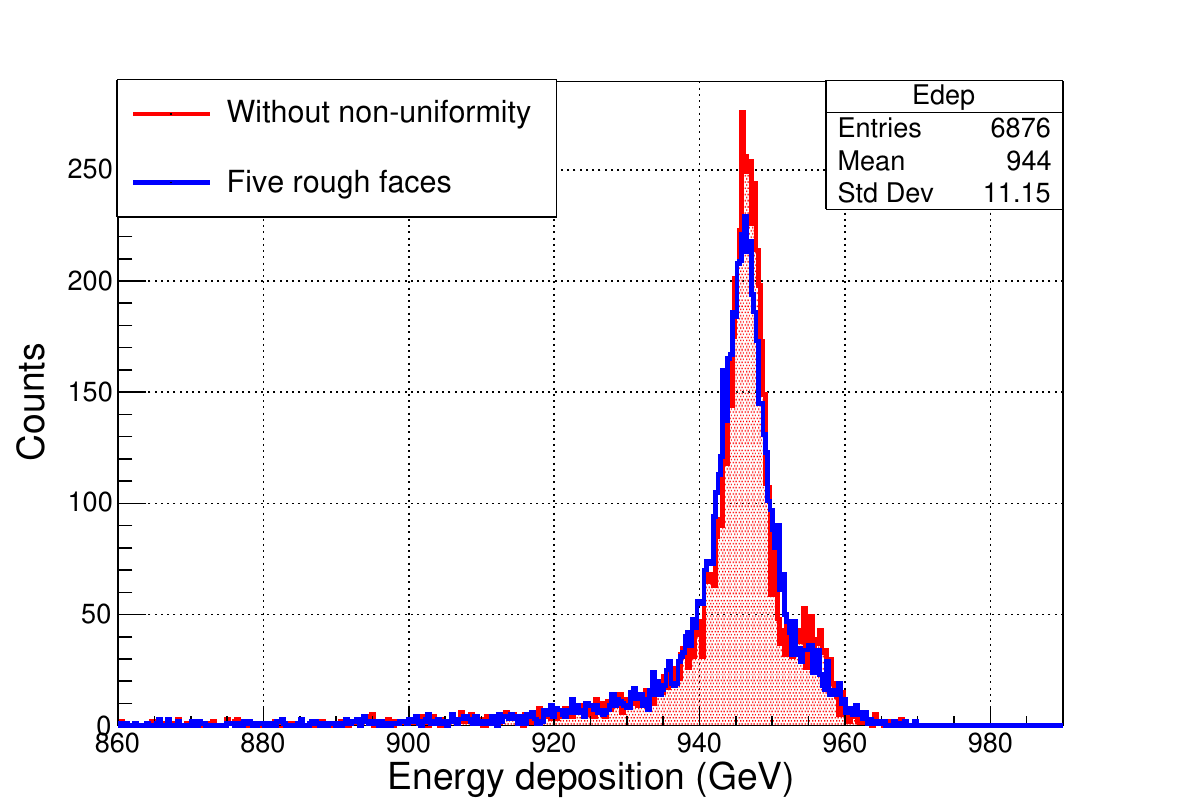}
\label{Fig3-4}
}
\caption{Reconstructed energy spectra of 1000 GeV electrons. A reconstructed energy spectrum without non-uniformity was also added for comparison.}\label{fig:1000gev}
\end{figure}

\begin{figure}[ht]
\centering
\includegraphics[width=4in]{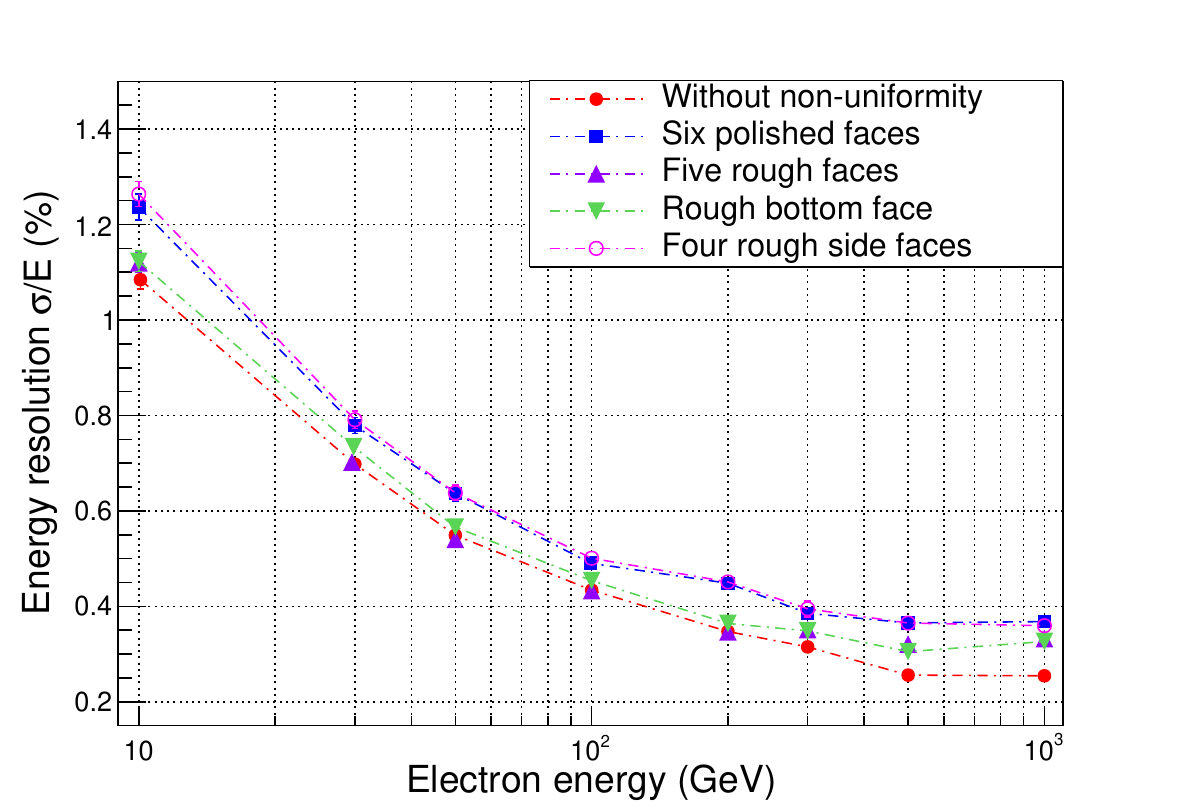}
\caption{The comparison of electron energy resolution among four surface roughness configurations with energies ranging from 10 to 1000 GeV.}\label{fig:electron}
\end{figure}

The reconstructed energy spectra of 1 TeV electrons with different surface roughness are showed in Figure \ref{fig:1000gev}. The electron energy resolution with energies ranging from 10 to 1000 GeV is summarized in Figure \ref{fig:electron}. The cube with five rough surfaces and the cube with rough bottom face exhibit much better energy resolution for electrons than the other two configurations.

Table \ref{tab3} summarizes the  $\sigma$ and $\gamma$  of the positional correlation factor distribution for different surface roughness. The $\sigma$ and $\gamma$ are generally consistent with the surface scanning results. For incident electron energies less than 1000 GeV, only low-range fibers are adopted, the cube with five rough surfaces and the cube with rough bottom face have smaller energy resolution resulting from smaller $\sigma$ or $\gamma$. When the incident electron energy is 1000 GeV, high-range fibers are starting to be utilized, the energy resolution of the four configurations becomes closer due to the cube with six polished faces or with four rough side faces has smaller $\gamma$ values on high-range fiber output .

\begin{table}[htbp]
\centering
\caption{Statistical results($\sigma$ and $\gamma$) of the positional correlation factor distribution for different surface roughness.}
\label{tab3}
\begin{tabular}{|cc|c|c|c|c|}
\hline
\multicolumn{2}{|c|}{Surface condition}  & \begin{tabular}[c]{@{}c@{}}Six polished\\ faces\end{tabular} & \begin{tabular}[c]{@{}c@{}}Rough bottom\\ face\end{tabular} & \begin{tabular}[c]{@{}c@{}}Four rough \\ side faces\end{tabular} & \begin{tabular}[c]{@{}c@{}}Five rough\\ faces\end{tabular} \\ \hline
\multicolumn{1}{|c|}{\multirow{2}{*}{\begin{tabular}[c]{@{}c@{}}Low range\\ fiber\end{tabular}}}  & Std Dev  & 0.043   & 0.036 & 0.044  & 0.040  \\ \cline{2-6} 
\multicolumn{1}{|c|}{}   & Skewness & -0.951  & -0.934   & -1.855   & 0.229   \\ \hline
\multicolumn{1}{|c|}{\multirow{2}{*}{\begin{tabular}[c]{@{}c@{}}High range\\ fiber\end{tabular}}} & Std Dev  & 0.056   & 0.052   & 0.056   & 0.054  \\ \cline{2-6} 
\multicolumn{1}{|c|}{}    & Skewness & 0.696 & 1.845   & -0.740    & 1.042   \\ \hline
\end{tabular}
\end{table}

\section{Conclusion}
This study investigates the combination of different surface roughness and reflective materials for the LYSO cubes. Results show that the cube with five rough faces and ESR coating produces the highest amplitude at the WLS fiber end, resulting in a roughly 44\% increase compared to a polished cube with PTFE tape wrapping. The non-uniformity of amplitude dependence on fired positions within the cube was measured, and the positional correlation factor was derived for the entire cube. To evaluate the effect on energy resolution, a simulation based on HERD CALO was conducted, which revealed that both the LYSO cube with five rough faces and the cube with rough bottom face exhibit much better energy resolution for electrons than the other two configurations. The smaller $\sigma$ and $\gamma$ of the position response distribution are beneficial for improving the energy resolution of the calorimeter.

The stopping power of LYSO for most X-rays in this measurement is less than 0.5 mm, thus it is necessary to measure the non-uniformity using MIP signals to verify the results of X-ray scanning.
This work provides guidance on selecting appropriate surface roughness and coating materials for LYSO, which are crucial for payload development. The final selection of the payload version requires a combination of factors including the amplitude of WLS fiber output, non-uniformity in position response, and adaptability to environmental conditions.

\acknowledgments

This work is supported by the National Natural Science Foundation of China, Grant No. 12027803, the International Partnership Program of Chinese Academy of Sciences, Grant No. 113111KYSB
20190020, the National Key R\&D Program of China, Grant No. 2021YFA0718401.

% Bibliography

%% [A] Recommended: using JHEP.bst file
\bibliographystyle{JHEP}
\bibliography{biblio.bib}

\end{document}